 \documentclass[final,1p,times,twocolumn]{elsarticle}

\usepackage{amssymb}
\journal{Journal of High Energy Astrophysics}

\begin{document}

\begin{frontmatter}



\title{High energy signatures of quasi-spherical accretion onto rotating, magnetized neutron star in the ejector-accretor intermediate state}


\author{W. Bednarek and P. Banasi\'nski}

\address{Department of Astrophysics, University of \L \'od\'z,
              ul. Pomorska 149/153, 90-236 \L \'od\'z, Poland}

\begin{abstract}
We consider a simple scenario for the accretion of matter onto a neutron star in order to understand processes in the inner pulsar magnetosphere during the transition stage between different accretion modes.
A simple quasi-spherical accretion process onto rotating, magnetized compact object is analysed in order to search 
for the radiative signatures which could appear during transition between ejecting and accreting modes.
It is argued that different accretion modes can be present in a single neutron star along different magnetic field lines for specific range of parameters characterising the pulsar (rotational period, surface magnetic 
field strength) and the density of surrounding medium. The radiation processes characteristic for the ejecting pulsar, i.e. curvature and synchrotron radiation produced by primary electrons in the pulsar outer gap, are expected to 
be modified by the presence of additional thermal radiation from the neutron star surface.
We predict that during the transition from the pure ejector to the pure accretor mode (or vice versa) an intermediate accretion state can be distinguished which is characterised by the $\gamma$-ray spectra of pulsars truncated below 
$\sim 1$ GeV due to the absorption of synchro/curvature spectrum produced in the pulsar gaps. 
\end{abstract}


\end{frontmatter}

% \linenumbers

\section{}
\label{}

\section{Introduction}

The importance of the accretion of matter onto compact objects as a key process for generation of energy around compact objects has been recognized since early 70-ties. In fact, three basic modes for the interaction of compact object with the surrounding matter have been distinguished, i.e. the ejector, accretor and propeller modes. These different models have been observed and investigated in the broad energy range. It is expected that neutron stars (NSs) can change their accretion modes with the evolution of their parameters (rotational period, magnetic field strength) or with the change of the parameters of the surrounding medium.
In fact, the external conditions around ejecting pulsars can change significantly when they enter from time to time dense clouds in the interstellar medium. The surrounding plasma, attracted by strong gravitational field of the neutron star,
can overcome the pressure of the pulsar wind initializing the transition from the ejecting to the accreting pulsar.
Similar transition process can also happen in the opposite direction when the accreting neutron star emerges from a  
dense cloud into a rare interstellar medium.
Recently, transition stages between the ejector and accretor modes have been discovered in a few millisecond pulsars within low mass binary systems  (PSR J1023+0038~\citep{ar09,pa14,st14}, IGR J18245-2452~\citep{pa13}, and XSS J12270-4859\citep{dm10,ro14}). Such transition from the ejecting stage (active radio pulsar) to the accreting stage resulted also in an enhanced $\gamma$-ray emission above $>100$ MeV~\citep{tll14,tam14,xi14}. 
In the case of sources mentioned above, the appearance of an accretion disk is well documented in one of the transition stages. Therefore, the accretion process occurs in a quite complicated way. A few different models have been recently proposed as  possible explanation of the transiting MSPs, see~\citep{ta14,ptl14,be15}.
For example, Bednarek~\citep{be15} argues that the accretion disk penetrates in the equatorial region of the rotating neutron star but the radiation processes typical for ejecting pulsars can still operate above the equatorial plane, e.g. as expected in terms of the slot gap model.

In the present paper we suggest that such transition stages can be also present in the case of
isolated pulsars which are free flying in the interstellar space. In the case of a pulsar immersed in 
a spherically-symmetric cloud, simple accretion geometry might also put new light on a more complicated processes occurring during transition stages in neutron stars accreting matter in the accretion disk geometry.
In this paper we consider the simplest possible case of a spherically-symmetric accretion onto rotating neutron star. We show that also in such geometrically simple accretion scenario the transition between the accretion mode and the radio pulsar mode could turn to the specific radiation features in $\gamma$-ray energies. We analyse the case of quasi-spherical accretion of matter onto magnetized, rotating millisecond pulsar which magnetic dipole is aligned with the pulsar rotational axis. We speculate that physical processes, typical for ejecting and accreting pulsars, could occur along different magnetic field lines during the transition stage in a single object. The effects on the $\gamma$-ray spectrum emitted by pulsars in such simple scenario 
are discussed. The quasi-spherical accretion scenario allows easier understanding of complicated physics of the accretion process onto rotating and magnetized NS. We conclude that different states of $\gamma$-ray emission might be also observed in the case of a quasi-spherical accretion of matter from dense interstellar cloud.

\section{Hybrid accretion modes onto magnetized NS}

\begin{figure}[t]
\vskip 6.7truecm
\includegraphics{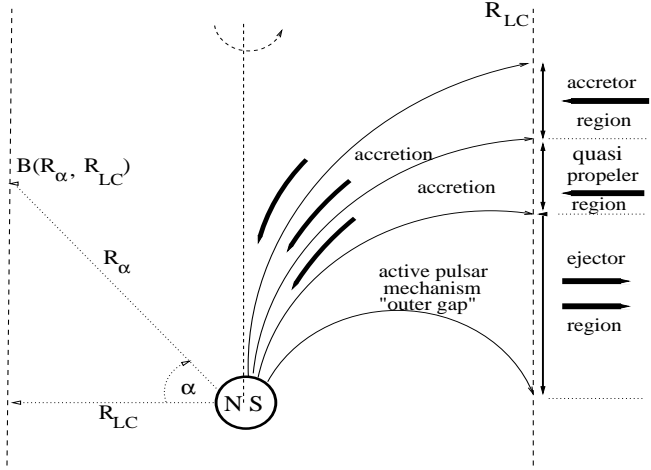}
\caption{Schematic representation of of the hybrid accretion scenario in which a part of the magnetosphere
is still in the ejector mode and a part of the pulsar magnetosphere is already able to accrete the matter distributed quasi-spherically. Since a part of the magnetosphere, close to the rotational plane of the pulsar, is free of matter, the outer gap acceleration model for the radiation processes~\citep{chr86} can still 
operate in this part of the magnetosphere which is connected to the magnetosphere outside the light cylinder radius
in the ejector stage. We define the critical angle, $\alpha_{\rm cr}$, as the angle $\alpha$ which separates 
a part of the pulsar magnetosphere in the ejector region from a part in the accretor region.
The angle $\alpha_{\rm cr}$ is determined by the condition that the magnetic field energy density, at 
the light cylinder ($R_{\rm LC}$) and at the distance $R_\alpha$ from the centre of the neutron star, 
is equal to the free fall kinetic energy of accreting matter (see Eq.~3).
Then, the pulsar magnetosphere is composed of two parts. The accretion occurs with almost free fall velocity
along the magnetic field lines which are connected to accretor region. The accretion occurs 
with the velocity significantly reduced from the free fall velocity, due to the partial balance of 
the gravitational force by the centrifugal force, along the magnetic field lines which lay in the region called quasi-propeller.}
\label{fig1}
\end{figure}

Neutron stars, with the mass of 1.4 of the Solar mass ($M_{\rm NS} = 1.4M_\odot$) and the radius of $R_{\rm NS} = 10$ km, produce strong gravitational field which effectively attracts background matter from the surrounding medium or 
the matter expelled from the stars in the form of stellar winds. 
In the simplest possible case, when the background matter has relatively low angular momentum in respect to the NS,
the matter falls onto the NS quasi-spherically. For example, such quasi-spherical accretion mode is
expected 
in the case of a strong isotropization of the matter after passing through the shock due to the movement of the NS.
The rate of an in-falling matter is described by the so called Bondi accretion rate~\citep{bh44,bo52}. It depends on the density of surrounding medium (a cloud), $n_{\rm cl} = 10^4n_4$ cm$^{-3}$, 
and on the relative velocity of the matter in respect to NS, 
$v = (v_{\rm NS}^2+v_{\rm T}^2)^{1/2}\approx 10^2v_2$ km s$^{-1}$,
where $v_{\rm NS} = 10^2v_2$ km s$^{-1}$ is the velocity of the NS, $v_{\rm T} = 3k_{\rm B}T/m_{\rm H}\approx 9.3T_4$ km s$^{-1}$ is the thermal velocity of the gas particles, $T = 10^4T_4$ K is the gas temperature, $k_{\rm B}$ is the Boltzmann constant, and $m_{\rm H}$ is the hydrogen atom mass.
The Bondi accretion rate can be expressed by,
\begin{eqnarray} 
\dot{M}_{\rm B} = 4\pi R_{\rm B}^2vn_{\rm cl}m_{\rm p}\approx 1.4\times 10^{13}n_4/v_{2}^{3}~~~{\rm g~s^{-1}}, 
\label{eq1}
\end{eqnarray}
\noindent
where the Bondi radius is $R_{\rm B} = 2GM_{\rm NS}/v^2\approx 2.7\times 10^{12}/v_2^2$ cm. 
In the case of accretion of matter onto rotating and magnetized NS, the fate of the matter is determined by the parameters of the NS (its period, magnetic field strength) and the surrounding matter (density and temperature). If the temperature of the accreting plasma is large than the accreting flow can be stopped by the magnetic field~\citep{al76}. This critical temperature is of the order of $\sim$(0.1-0.3)$T_{\rm ff}$, where $T_{\rm ff} = GM_{\rm NS}m/k_{\rm B}R\approx 1.6\times 10^{12}/R_6$ K is the free fall equipartition temperature and $R = 10^6R_6$ cm is the distance from the NS. We assume that the plasma temperature is low enough to allow spherical accretion of matter onto NS. 
Three stages for interaction of NS magnetosphere with the surrounding matter can be defined, ejector, propeller, and accretor modes (see~\citep{is75}, for more recent review see~\citep{li92} and references therein). Specific modes occur for specific parameters of the NS and surrounding matter. However, the parameters of the surrounding medium in which NS is immersed can change significantly. Therefore, transitions between different modes should be also observed for specific values of the parameters 
defining accretion scenario. 
The question appears how the transition from the ejecting to the accreting pulsar occurs ? Which parts of the pulsar inner magnetosphere are at first penetrated by the matter, equatorial or polar ? We speculate that at first the matter penetrates the polar regions of the pulsar inner magnetosphere since the magnetic field at the light cylinder in those regions is weaker than at the light cylinder close to the rotational plane of the pulsar.
We argue that during such transition stages different modes of accretion can co-exist 
in a single object. We are mainly interested in transitions between ejector and accretor modes which we call 
intermediate accretion mode. Such intermediate mode is possible since  the conditions at different parts of the light cylinder of the rotating neutron star with the dipole structure of the magnetic field can differ significantly. In order to easier understand the basic features of such intermediate state, we assume that 
the magnetic moment of the 
NS is co-aligned with the rotational axis of NS (see Fig.~1 for schematic representation of the geometry).
We show that for specific accretion rates the magnetic field energy density at the regions of the light cylinder, which are far away from the rotational plane of the pulsar, can become 
lower than the energy density of an accreting plasma thus allowing penetration of matter towards the inner pulsar
magnetosphere. On the other hand, the magnetic field energy density can be still larger than the kinetic energy density
of the matter at regions of the light cylinder which are close to the rotational plane. As a result, the accretion
of plasma can still occur along the magnetic field lines which cross the light cylinder far away from the
rotational plane, i.e. for large values of the angle $\alpha$ (see Fig.~1), as expected in the accretor and/or
propeller mode of accretion. In contrast, the plasma is still effectively ejected in the region close to the
rotational plane of the pulsar which still operates in the  ejector mode. We expect that for some range of the
accretion rates, different stages (accretor and ejector) can be present in  a single NS immersed in the cloud of
ionized matter. The details of such hybrid scenario and expected radiation processes during the transition stage
between different modes are discussed in this paper.

\subsection{Accreting versus ejecting magnetosphere}

Now we evaluate which part of the magnetosphere is occupied by different accretion modes for different parameters
of the NS and surrounding medium. We assume that the NS is immersed in the isotropic cloud of ionized matter. The accretion rate in the case of quasi-spherical scenario will be described by the Bondi accretion rate, 
$\dot{M}_{\rm B}$ (see Eq.~1).
Then, the density of matter in different parts of the light cylinder depends on the distance from the NS centre,
$R_\alpha$, as,
\begin{eqnarray}
\rho = \dot{M}_{\rm B}/(4\pi R_{\alpha}^2V_{\rm ff})~~~{\rm g~cm^{-3}},
\label{eq2}
\end{eqnarray}
\noindent
where $V_{\rm ff} = \sqrt{2GM_{\rm NS}/R_\alpha}$ is the free fall velocity of accreting matter, $R_\alpha =
R_{\rm LC}/\cos\alpha$ is the distance to the light cylinder at the angle $\alpha$ (see Fig.~1), $R_{\rm LC} = cP/2\pi\approx  4.77\times 10^6P_{\rm ms}$ cm is the light cylinder radius, $P = 10^{-3}P_{\rm ms}$ s is the pulsar period, $c$ is the velocity of light, and $G$ is the gravitational constant. On the other hand, also the energy density of the magnetic field at different parts of the light cylinder 
is a function of the angle $\alpha$.
The condition which separates the ejecting part of the NS magnetosphere and the accreting part of the magnetosphere 
can be obtained by comparing the magnetic energy density with the kinetic
energy density of accreting matter at the light cylinder radius~\citep{po73,lpp73,do73}. These energy densities depend on the angle $\alpha$ according to,
\begin{eqnarray}
{{B^2(R_\alpha, R_{\rm LC})}\over{8\pi}} = {{\rho V_{\rm ff}^2}\over{2}} = 
{{GM_{\rm NS}\rho}\over{R_\alpha}},
\label{eq3}
\end{eqnarray}
\noindent
where the magnetic field at the light cylinder at the distance $R_\alpha$ from the neutron star is given by,
\begin{eqnarray}
B(\alpha, R_{\rm LC}) = \sqrt{(B^2_x(\alpha, R_{\rm LC} +B^2_z(\alpha, R_{\rm LC})},
\label{eq4}
\end{eqnarray}
\noindent
with the components of the magnetic field strength,
\begin{eqnarray}
B_x(\alpha, R_{\rm LC}) = 3\mu_{\rm NS}\sin\alpha(\cos\alpha)^4/R_{\rm LC}^3,
\label{eq5}
\end{eqnarray}
\noindent
and
\begin{eqnarray}
B_z(\alpha, R_{\rm LC}) = \mu_{\rm NS}[1 - 3(\sin\alpha)^2](\cos\alpha)^3/R_{\rm LC}^3.
\label{eq6}
\end{eqnarray}
\noindent
$\mu_{\rm NS} = B_{\rm NS}R_{\rm NS}^3\approx 10^{26}B_8$ G~cm$^3$ is the magnetic moment of the NS and $B_{\rm NS} = 10^8B_8$ G is its surface magnetic field strength.

\begin{figure}[t]
\vskip 5.5truecm
\includegraphics{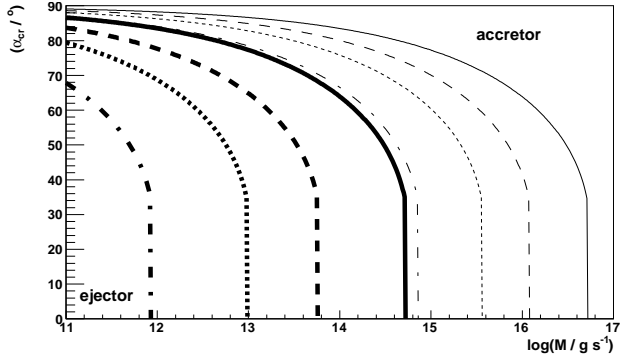}
\caption{The value of the critical angle as a function of different accretion rates (calculated for the case of quasi-spherical accretion) for a few selected NS rotational periods, $P_{\rm ms} = 1.6$ ms (solid curve), 
3 ms (dashed), 5 ms (dotted), and 10 ms (dot-dashed). The ejector mode is expected for the low accretion rates 
and small angles $\alpha$ (below thick curves). The accretor mode is expected for the angle $\alpha_{\rm cr}$ above the thick curves. The thin curves separate the quasi-propeller and accretor modes of accretion onto 
the NS surface. The magnetic moment of the neutron star is fixed on $10^{26}$ Gs cm$^3$.}
\label{fig2}
\end{figure}

We determine the critical value of the angle $\alpha_{\rm cr}$ for which energy density of 
the magnetic field is equal to 
the kinetic energy density of the accreting matter by solving Eq.~3. This angle separates the accreting part of the
magnetosphere from the ejecting part of the magnetosphere as shown in Fig.~1. Since the angle $\alpha$ is involved 
in Eq.~3 in a complicated way, we solve this equation numerically. The results of calculations are shown in Fig.~2 
for a few
selected values of the pulsar period and as a function of the accretion rate of matter onto NS. Note that, for 
specific accretion rate the value of the angle $\alpha_{\rm cr}$ drastically drops to zero. 
For larger accretion rates the ejecting part of pulsar magnetosphere disappears completely and the high 
energy radiation processes, characteristic for ejecting radio pulsars, are quenched. The parameter space 
(accretion rate versus pulsar period), at which such "pure accretor" mode is expected, is shown in Fig.~3 
for the case of the millisecond pulsars and also for the classical pulsars. 
The specific curves in this figure separate the "pure accretor" from the "partial ejector" mode for a few 
selected values of the dipole magnetic moment of the neutron star, corresponding to the range of surface magnetic field strengths $10^8-10^{10}$ Gs (in the case of millisecond pulsars, upper figure) and $10^{12}-10^{14}$ Gs (in the case of classical pulsars, bottom figure). The hybrid accretion modes are expected in the case of the millisecond pulsars which are able to accrete the matter close to the curves in Fig.~3. These curves separate partial ejector from pure accretor modes. Note also that, these values of the accretion rates correspond to specific thermal X-ray luminosities of the pulsars in the accretor modes (see Eq.~10). For typical soft X-ray luminosities in the range $10^{31}-10^{38}$ erg s$^{-1}$, the accretion rates are expected to be in the range 
$\sim 10^{11}-10^{18}$ g s$^{-1}$. Therefore, the hybrid accretion mode can be characterised by  
the specific pulsed X-ray emission from the surface of the NS.  

\begin{figure}[t]
\vskip 9.truecm
\includegraphics{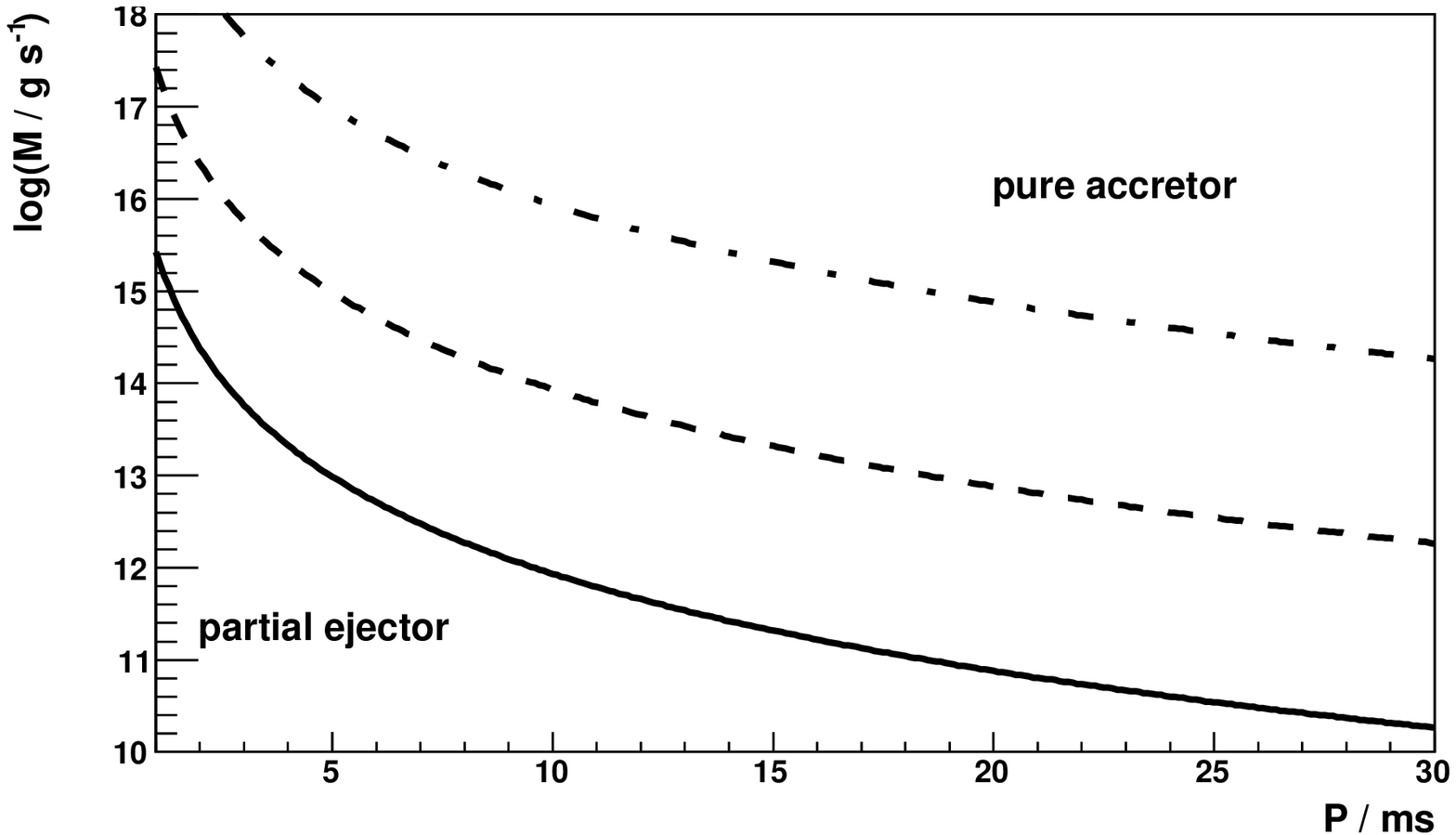}
\includegraphics{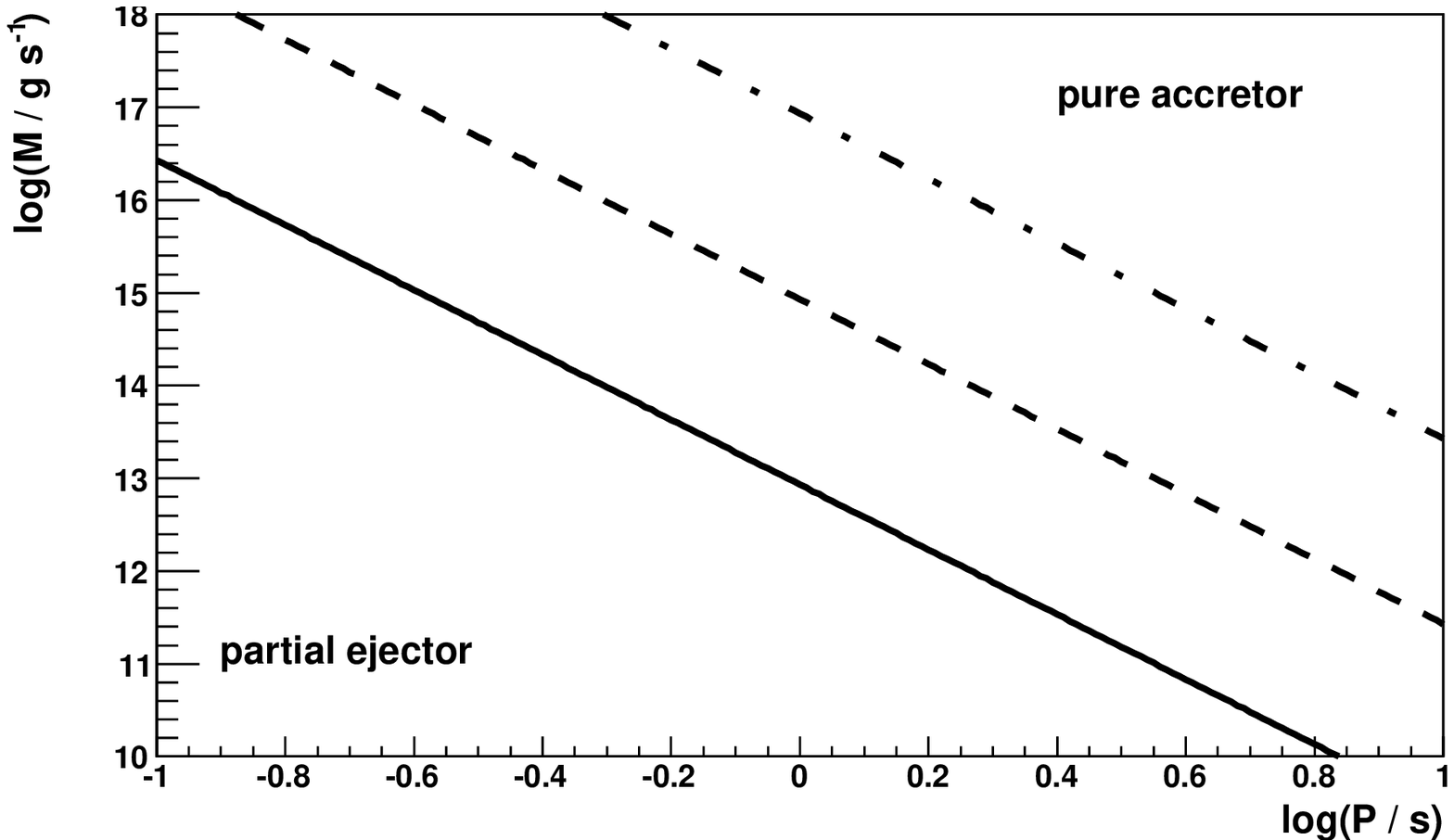}
\caption{The parameter space (the pulsar period versus the Bondi accretion rate) for the partial ejector mode and only (pure) accretor mode for the case of millisecond pulsars
(upper panel) and classical pulsars (bottom panel). The magnetic moment of the NS is equal to 
$\mu_{\rm NS} = 10^{26}$ Gs cm$^3$ (solid curve), $10^{27}$ Gs cm$^3$ (dashed curve), and $10^{28}$ Gs cm$^3$ 
(dot-dashed curve) in the upper panel and for $10^{30}$ Gs cm$^3$ (solid), $10^{31}$ Gs cm$^3$ (dashed), and 
$10^{32}$ Gs cm$^3$ (dot-dashed) in the bottom panel.}
\label{fig3}
\end{figure}

The accretor and quasi-propeller regions in the inner pulsar magnetosphere are separated by the condition that the 
Alfven radius, for the matter penetrating the inner pulsar magnetosphere, is comparable to the co-rotation radius. 
The co-rotation radius is calculated by comparing the centrifugal force of the matter, which is attached to the magnetic field lines, with the gravitational force, i.e. $\rho v_{\rm r}^2/X = GM_{\rm NS}\rho \cos\alpha/R_\alpha^2$, where $v_{\rm r} = 2\pi X/P$ is the rotational velocity of the magnetic field lines, $R_\alpha = X/\cos\alpha$, and X is the distance from rotational axis (see Fig.~1). This condition gives us the co-rotation radius as a function of the angle $\alpha$,
\begin{eqnarray}
X_{\rm cor} \approx 1.5\times 10^6P_{\rm ms}^{2/3}\cos\alpha~~~{\rm cm}.
\label{eq7}
\end{eqnarray}
\noindent
The Alfven radius is obtained by comparing the energy density of the in-falling plasma (equal to its gravitational energy) with the energy density of the magnetic field, $U_{\rm B} = U_{\rm kin} = U_{\rm grav}$, 
\begin{eqnarray}
GM_{\rm NS}\rho/R_{\alpha} = [B_{\rm x}^2(R_{\rm \alpha}) + B_{\rm z}^2(R_{\rm \alpha})]/8\pi,
\label{eq8}
\end{eqnarray}
\noindent
where $B_{\rm x}(R_{\rm \alpha})$ and $B_{\rm z}(R_{\rm \alpha})$ are the components of the magnetic field at the distance $R_{\rm \alpha}$ (see Eq.~5 and 6).
The above equation allows us to determine the Alfven radius, $x_{\rm A}$, as a function of the angle $\alpha$,
where $x_{\rm A} = R_\alpha\cos\alpha$. The accretor mode is separated from the quasi-propeller mode by the condition, $x_{\rm cor} = x_{\rm A}$. This condition is reached for specific value of the angle $\alpha$.
We have solved Eq.~7 and Eq.~8 numerically in order to determine the angle $\alpha_{\rm p/a}$, which separates the quasi-propeller region from the accretor region in the inner pulsar magnetosphere. We expect that in the case of quasi-propeller mode the accretion of matter also occurs onto the surface of NS since the centrifugal force balances only the component of the gravitational force which is perpendicular to the rotational axis of the NS
but not the component along the magnetic field lines. However, in such case the accretion process, along the magnetic field lines which touches the light cylinder in the range of angles $\alpha_{\rm cr}$ and $\alpha_{\rm p/a}$, occurs with lower in-fall velocity than the free fall velocity.

\subsection{Radiation of accreting matter}

The amount of matter which is expected to reach the magnetic pole on the NS surface is a part of the Bondi accretion rate,
\begin{eqnarray}
\dot{M}_{\rm acc} = \dot{M}_{\rm B}[1 - \cos(90^{\rm o} - \alpha_{\rm cr})]/2 = a_{\rm cr}\dot{M}_{\rm B}.
\label{eq9}
\end{eqnarray} 
\noindent
The gravitational energy released on the NS surface due to accretion of this matter is,
\begin{eqnarray}
L_{\rm acc} = G\dot{M}_{\rm acc} M_{\rm NS}/R_{\rm NS}\approx  
1.9\times 10^{33}M_{13}a_{\rm cr}~~~{\rm erg~s^{-1}},
\label{eq10}
\end{eqnarray}
\noindent
where the accretion rate is in units of $\dot{M}_{\rm acc} = 10^{13}M_{13}$ g s$^{-1}$, and the accretion luminosity will be scaled by $L_{\rm acc} = 10^{33}L_{33}$ erg s$^{-1}$. 

We assume that this accretion energy is thermalized on the NS surface.
The characteristic temperature of the polar cap region on the NS surface can be estimated for specific 
radius of the polar cap, $R_{\rm cap}$. This radius is defined by the critical magnetic field line which touches the light
cylinder radius at the critical angle $\alpha_{\rm cr}$. The dipole magnetic field lines 
(in the polar coordinates) are well described by the equation $R = C\sin^2\theta$~\citep{fkr85}. The constant $C$ can be evaluated from the condition that the magnetic field line passes through the light cylinder radius at the angle $\alpha_{\rm cr}$. Then, $C = R_{\rm LC}/[\sin (90^o-\alpha_{\rm cr})]^{-3}$. The radius of the polar cap is given by, 
\begin{eqnarray}
R_{\rm cap} = {{R_{\rm NS}^{3/2}(\cos\alpha_{\rm cr})^{3/2}}\over{R_{\rm LC}^{1/2}}}\approx
{{4.6\times 10^5(\cos\alpha_{\rm cr})^{3/2}}\over{P_{\rm ms}^{1/2}}}~~~{\rm cm}.
\label{eq11}
\end{eqnarray}
\noindent
It defines the surface area of the polar cap $S_{\rm cap} = \pi R_{\rm NS}^3(\cos\alpha_{\rm cr})^3/R_{\rm LC} \approx
6.6\times 10^{11}(\cos\alpha_{\rm cr})^3/P_{\rm ms}$ cm$^2$.
Then, the temperature of the polar cap is estimated on,  
\begin{eqnarray}
T_{\rm cap} = \left({{L_{\rm acc}}\over{\sigma_{\rm SB}S_{\rm cap}}}\right)^{1/4}\approx 
2.3\times 10^6\left({{{L_{33}P_{\rm ms}}}\over{\cos^3\alpha_{\rm cr}}}\right)^{1/4}~~~{\rm K.}
\label{eq12}
\end{eqnarray}
\noindent
We show in Fig.~4 the temperature of the hot polar cap region on the NS surface as a function of the Bondi accretion rate for a few selected values of the magnetic moment of the NS. The surface temperature increases for smaller accretion rates but the area of the hot polar cap decreases. The temperature will not increase infinitly since the accretion process cannot
occur inside the cone around the rotational axis defined by the condition $R_\alpha < R_{\rm B}$. In fact, the hot polar cap
has a hollow inside due to this condition. This condition also defines the maximum value of the angle $\alpha_{\rm max}$ above which the accretion process onto neutron star surface cannot occur i.e., $\cos\alpha_{\rm max} = R_{\rm LC}/R_{\rm B}$.

\begin{figure}[t]
\vskip 9.truecm
\includegraphics{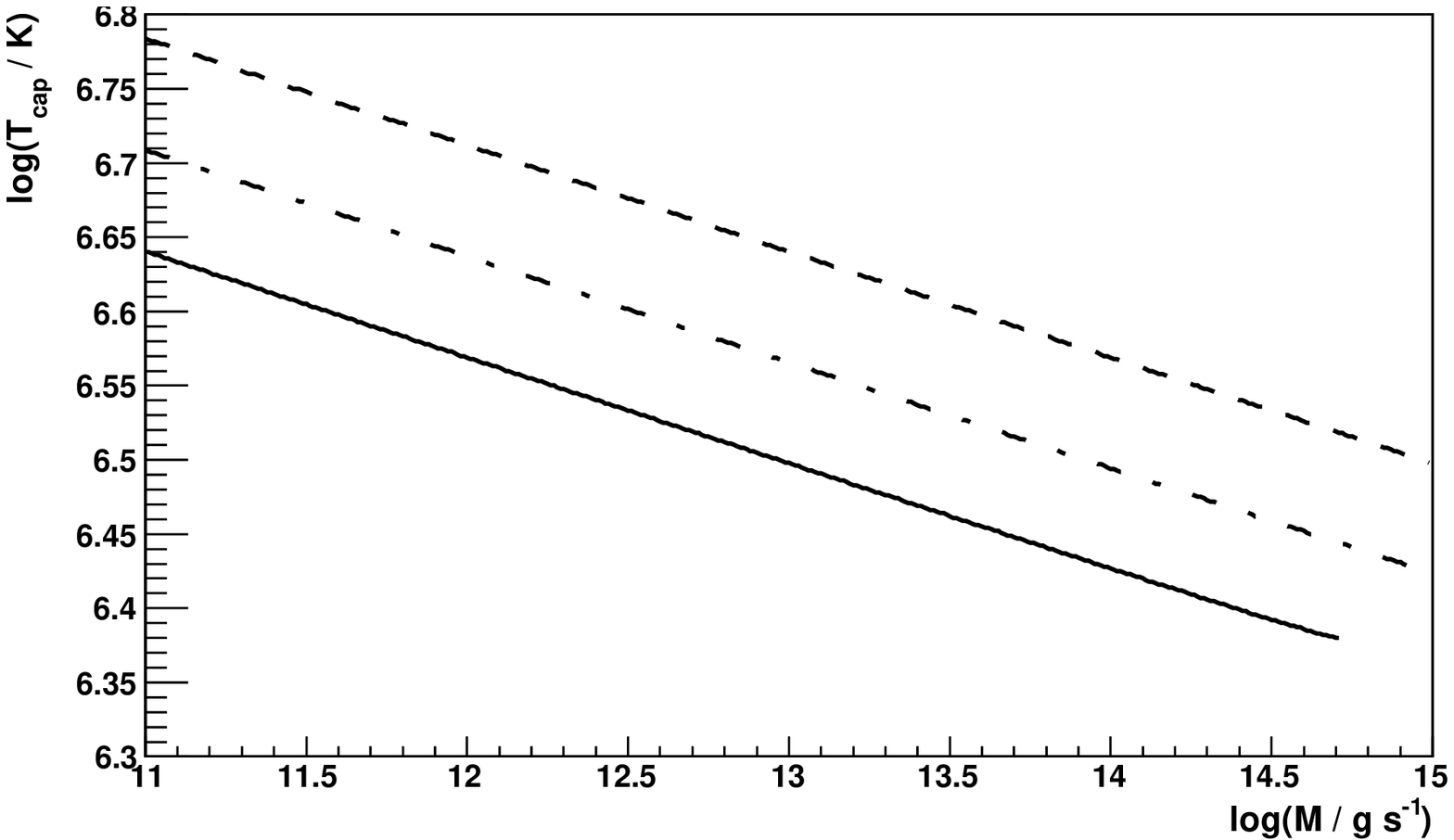}
\includegraphics{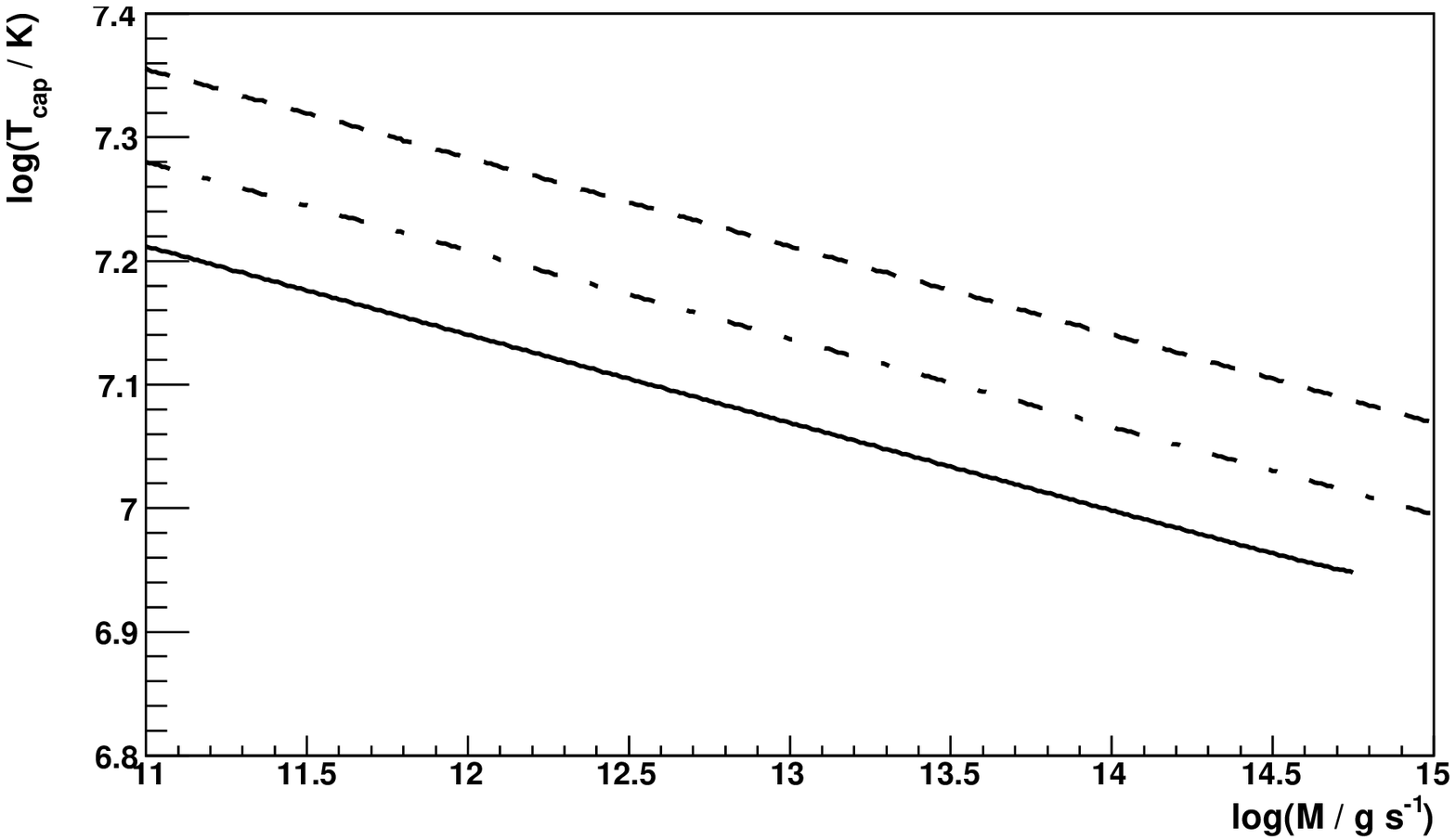}
\caption{The surface temperature of the hot polar cap region as a function of the Bondi accretion rate for selected values of the magnetic moment of the neutron star fixed on $10^{26}$ Gs cm$^3$ (solid curve), $3\times 10^{26}$ Gs cm$^3$ (dashed) and $10^{27}$ Gs cm$^3$ (dot-dashed) and the millisecond pulsar with the period equal to 1.6 ms (upper panel). The case of the classical pulsar with
the period of 300 ms and the the magnetic moment of the neutron star fixed on $10^{30}$ Gs cm$^3$ (solid curve), $3\times 10^{30}$ Gs cm$^3$ (dashed) and $10^{31}$ Gs cm$^3$ are shown in the bottom panel.}
\label{fig4}
\end{figure}

The density of thermal photons, emitted from this hot polar region, at the distance of the light cylinder radius (i.e. the approximate distance of the outer gap region) is estimated on,
\begin{eqnarray}
\rho_{\rm th}\approx {{L_{\rm acc}}\over{c S_{\rm cap}}}\left({{R_{\rm cap}}\over{R_{\rm LC}}}\right)^2 
\approx 4.7\times 10^8  L_{33}/P_{\rm ms}^2~~~{\rm erg~cm^{-3}},
\label{eq13}
\end{eqnarray}
\noindent
where $\sigma_{\rm SB}$ is the Stefan-Boltzmann constant. We estimate the
mean free path for leptons on the IC scattering of this thermal radiation field (in the Thomson regime) on,
\begin{eqnarray}
\lambda_{\rm IC} = 3m_{\rm e}c^2\gamma_{\rm e}/(4\sigma_{\rm T}\rho_{\rm th}\gamma_{\rm e}^2)\approx 
2\times 10^9P_{\rm ms}^2/(L_{33}\gamma_{\rm e})~~~{\rm cm}, 
\label{eq14}
\end{eqnarray}
\noindent
where $\sigma_{\rm T}$ is the Thomson cross section, and $\gamma_{\rm e}$ is the electron Lorentz factor.
The comparison of this mean free path with the characteristic distance scale for their propagation within the inner magnetosphere, i.e. comparable to 
the light cylinder radius, allows us to estimate  the energies of leptons,  
$E_{\rm e}\approx 200 P_{\rm ms}/L_{33}$ MeV, for which they lose efficiently energy on the IC process. For some parameters, leptons with such energies can scatter thermal X-rays from the NS surface close to the T regime since the IC scattering at the transition between the Thomson (T) and the Klein-Nishina (K-N) regimes occurs for $E_{\rm T/KN}\approx m^2_{\rm e}c^4/(3k_{\rm B}T_{\rm cap})\approx 
320[(\cos\alpha_{\rm cr})^3/L_{33}P_{\rm ms}]^{1/4}$ MeV. The example $\gamma$-ray spectra, produced by the secondary leptons in the outer gap region, are considered in the next section.

\section{$\gamma$-ray emission in the intermediate state}

The high energy $\gamma$-ray production in the inner magnetospheres from over hundred pulsars have been 
well documented in the recent years~\citep{ab13}. Although, the details of these radiation 
and particle acceleration processes are still not exactly known, it is obvious that these processes
occur in the pulsar magnetosphere which operates in so called ejector mode, during which relativistic plasma 
is ejected through the pulsar light cylinder radius and the external matter is not able to enter the 
inner pulsar magnetosphere. As we argue above at some intermediate state the ejector mode can be only active 
in a part of the pulsar magnetosphere and the rest of the magnetosphere (defined by different magnetic 
field lines touching the light cylinder radius) is still able to accrete the matter from the surrounding medium.
We expect that in such intermediate state the radiation processes in the pulsar magnetosphere should become 
more complicated. Apart from 
the curvature (and synchrotron) radiation of very energetic primary electrons accelerated in the outer gap
(with the Lorentz factors $\gamma_{\rm e}\sim 10^7$), expected to be responsible for the 
pulsed $\gamma$-ray spectrum, a population of low energy secondary leptons (which appears at the magnetic 
field lines above the outer gap as a result of partial absorption of curvature $\gamma$-rays) can produce additional component in the $\gamma$-ray spectrum. This component is due to the comptonization of the thermal
radiation from the NS surface produced by accreting matter along the magnetic field lines above the angle 
$\alpha_{\rm cr}$. It is rather difficult to consider in detail this additional emission due to the lack of detailed knowledge on the spectrum of secondary leptons. Thermal emission from the polar cap can be also responsible for
modification of the synchro/curvature $\gamma$-ray spectrum produced by primary electrons due to the absorption of
multi-GeV $\gamma$-rays. The modification of the synchro/curvature primary emission is considered below in a simple absorption scenario.

We assume that the outer gap model~\citep{chr86}  can still be active in the hybrid accretion scenario
discussed above. We calculate the expected curvature radiation spectrum in the outer gap model in order
to have an idea about possible conditions at which such additional spectral components could be also observed.
We apply the simple model for the acceleration of leptons (and their radiation) in the outer gap which details 
are summarized in~\citep{ng14}. In this model the primary 
electrons are accelerated in the electric field of the outer gap with the electric field strength,
\begin{eqnarray}
E_\parallel\sim f_{\rm gap}B_{\rm LC}R_{\rm LC}/R_{\rm cur}\sim 6\times 10^8f_{\rm g}B_8P_{\rm ms}^{-5/2} 
~~~{\rm V~cm^{-1}}, 
\label{eq15}
\end{eqnarray}
\noindent
where $R_{\rm cur}\approx \sqrt{R_{\rm LC}R_{\rm NS}}\approx  2.2\times 10^6 P_{\rm ms}^{1/2}$ cm, 
$B_{\rm LC} = B_{\rm NS}(R_{\rm NS}/R_{\rm LC})^3$ is the magnetic field strength at the light cylinder, 
and $f_{\rm gap}$ is the gap thickness typically of the order of $\sim 0.3$ for the millisecond pulsars~\citep{ta12}.
The $\gamma$-ray luminosity produced by primary electrons in the curvature process is,
\begin{eqnarray}
L_\gamma\approx f_{\rm gap}^3L_{\rm P}\approx  
3\times 10^{35}f_{\rm gap}^3B_8^2P_{\rm ms}^{-4}~~~{\rm erg~s^{-1}}.
\label{eq16}
\end{eqnarray}
\noindent
where $P_{\rm P}$ is the rotational energy loss rate of the pulsar.
Primary electrons accelerated in the outer gap produce observed pulsed emission in the curvature radiation 
process with the characteristic energies 
$E_{\rm cur} = 3hc\gamma_{\rm e}^3/4\pi R_{\rm cur}\approx 3\times 10^{-14}\gamma_{\rm e}^3/R_{\rm cur}$ GeV.
The energy loss rate of these electrons on the curvature process is $\dot{E}_{\rm cur} = 2e^2\gamma_{\rm e}^4/3R_{\rm cur}^2\approx 9.6\times 10^{-8}\gamma_{\rm e}^4/R_{\rm cur}^2$ eV cm$^{-1}$. The comparison of the acceleration efficiency, $eE_\parallel$, with the curvature energy loss rate allows us to estimate the equilibrium 
Lorentz factor of primary electrons in the outer gap
$\gamma_{\rm eq}\approx 1.3\times 10^7(f_{\rm g}B_8P_{\rm ms}^{-3/2})^{1/4}$. 
These electrons lose energy on characteristic distance scale, 
$\lambda_{\rm cur} = m_{\rm e}c^3\gamma_{\rm e}/\dot{E}_{\rm cur}$, which is, 
\begin{eqnarray}
\lambda_{\rm cur}\approx 5.2\times 10^{12}R_{\rm cur}^2/\gamma_{\rm e}^3\approx
1.2\times 10^5(f_{\rm g}B_8)^{-3/4}P_{\rm ms}^{17/8}~~~{\rm cm}, 
\label{eq17}
\end{eqnarray}
\noindent
for electrons with the equilibrium Lorentz factors. This is about an order of magnitude less than the light cylinder radius. So then, the process is efficient for electrons with the equilibrium energies.
We estimate the characteristic energies of curvature $\gamma$-rays produced by the primary electrons in the outer gap for 
the some example parameters of the pulsar, $B_{\rm NS} = 3\times 10^8$ G and $P = 3$ ms and the gap 
thickness $f_{\rm g} = 0.3$. The equilibrium Lorentz factor of electrons for these parameters is equal to 
$\gamma_{\rm eq}\approx 8.4\times 10^6$ and the characteristic energy of the curvature photons is
$E_{\rm cur}\approx 8.7$ GeV. However, curvature $\gamma$-rays with such energies might be efficiently absorbed in the thermal radiation from the polar cap. The absorption becomes important for $\gamma$-rays with energies,
\begin{eqnarray}
E_\gamma = 2 (m_{\rm e}c^2)^2/\varepsilon_{\rm th}\approx 840\left({{{L_{33}P_{\rm ms}}}\over{\cos^3\alpha_{\rm cr}}}\right)^{-1/4}
~~~{\rm MeV}.
\label{eq17b}
\end{eqnarray}
\noindent
where $\varepsilon_{\rm th} = 3k_{\rm B}T_{\rm cap}$, and $k_{\rm B}$ is the Boltzmann constant.
We estimate whether such absorption process of $\gamma$-rays is efficient by comparing the $\gamma$-ray mean free path, $\lambda_{\gamma\gamma}$  with the characteristic distance scale of the interaction process given by the light cylinder radius.
The mean free path can be estimated on,
\begin{eqnarray}
\lambda_{\gamma\gamma} = (n_{\rm ph}\sigma_{\gamma\gamma})^{-1}\approx 2\times 10^6 ({{{L_{33}P_{\rm ms}}/\cos^3\alpha_{\rm cr}}})^{-3/4}(\cos\alpha_{\rm cr}/P_{\rm ms})^{-3}~~~{\rm cm}
\label{eq17c}
\end{eqnarray}
\noindent
where $n_{\rm ph}\approx 380 (T_{\rm cap}/2.7)^3(R_{\rm cap}/R_{\rm LC})^2\approx 2.3\times 10^{18} ({{{L_{33}P_{\rm ms}}/\cos^3\alpha_{\rm cr}}})^{3/4}(\cos\alpha_{\rm cr}/P_{\rm ms})^3$ ph. cm$^{-3}$ and $\sigma_{\gamma\gamma}$ is the cross section for
$e^\pm$ pair production in $\gamma-\gamma$ collision.
This mean free path is shorter than the light cylinder radius for the pulsars with periods,
\begin{eqnarray}
P_{\rm ms}\approx 2(L_{33}\cos\alpha_{\rm cr})^{3/4}.
\label{eq17d}
\end{eqnarray}
\noindent
We conclude that in the case of short period pulsars and large accretion luminosities the curvature $\gamma$-rays with $\sim$1 GeV energies, produced in the outer gap by primary leptons, should be efficiently absorbed in the thermal radiation from the hot polar cap. Therefore, the $\gamma$-ray spectra in the transition state should be truncated at about $\sim$1 GeV in contrast to the
$\gamma$-ray spectra from observed MSPs which extend through the GeV energy range.

\section{Conclusion}

We argue that in the case of quasi-spherical accretion process the pulsar magnetosphere separates on two parts, the accreting part in the polar regions and the ejecting part in the equator regions. 
Both parts of the magnetosphere can be active in this same pulsar provided that
it is close to the transition from the ejector to the accretor stage (or vice versa). 
The duration of such transition state depends on the stability of medium in which MSP is immersed. In the case of considered MSP within a dense cloud it is determined by the homogeneity of the cloud. 
In such case the duration of the transition state is of the order of the dimension of the clamps within the cloud divided by the velocity of the pulsar. On the longer time scale the transition state can be also influenced by evolution of the parameters of the pulsar. Proposed hybrid magnetosphere is  
caused by different conditions in the pulsar magnetosphere at the light cylinder region at different distance from 
the rotational plane. For simplicity we analysed the case of a simple model for the neutron star which is 
an aligned magnetic rotator immersed in a quasi-spherical cloud of 
ionized matter. It is shown that for some parameters of the NS and the Bondi accretion rate of matter onto the NS,
a part of the magnetosphere close to the rotational plane is expected to be in the ejector mode and a part of 
the magnetosphere farther from the plane is already in the accretor mode (see Fig.~3). 
The accreting matter falls onto the small part of the NS surface creating a hot spot with 
the characteristic temperature of the order of a few million K.
This strong radiation creates additional background in the inner pulsar magnetosphere which has to be taken 
into account when analysing the radiation processes.

We propose that the synchro/curvature $\gamma$-ray spectrum produced by primary electrons in the inner magnetosphere 
during such intermediate state, should be truncated at $\sim 1$ GeV due to its absorption in the thermal radiation from 
the polar cap. Therefore the $\gamma$-ray spectra from the pulsars in the intermediate state should differ significantly 
from the pulsars in the ejecting state. In fact, the presence of additional strong radiation field in the pulsar gap region
should significantly modify also the gap acceleration mechanism and so curvature emission. However, it is difficult
to consider in detail acceleration and radiation processes in such additional conditions. Therefore we leave their detailed discussion to the future investigations.

Up to now none of observed isolated MSPs shows the change of the $\gamma$-ray emission state expected in this model.
The reason may be that none of these MSPs fulfil the conditions for the proposed state to occur. We have shown that
the MSP should be immersed in relatively dense environment. Moreover, The accretion rate should be not to large, since then the MSP moves to pure accretor state, but also not to low, since then the thermal luminosity from the neutron star surface
due to the accretion process is too weak to influence radiation processes in the inner pulsar magnetosphere.

Here we considered only the simple case of aligned (or almost aligned) rotator model for the pulsar 
magnetosphere in order to show that such two-component, accreting/ejecting magnetosphere of the pulsar might 
be observed. In the case of highly oblique rotators, the matter accreting through the light cylinder radius at 
large distances from the rotational plane may have problems with reaching the NS surface due to 
the geometry of the magnetic field lines. The magnetic field strength at the light cylinder will change
with the pulsar period. Therefore the conditions for accretion of matter onto NS will also periodically change.
As a result, the matter is expected to fall onto the NS surface in the form of bunches only when the condition given by Eq.~3 is fulfilled. Such transitional accretion, modulated with the period of the pulsar, does not need to be in phase with the
$\gamma$-ray emission from the parts of the magnetosphere near the light cylinder. However, the cumulative accretion rate 
is expected to be reduced in the case of oblique rotator. This will influence the thermal radiation field from NS surface and also eventual absorption of curvature/synchrotron radiation from the pulsar gaps.  
Therefore, we conclude that such characteristic truncated at $\sim 1$ GeV gamma-ray spectra are likely to be present only in the case of magnetic rotators which magnetic axis is nearly aligned to the rotational axis of the neutron star.

\section*{Acknowledgements}
We would like to thank the Referee for useful comments.
This work is supported by the grant through the Polish Narodowe Centrum Nauki No. 2011/01/B/ST9/00411.

\end{document}